\input{aipcheck}

\documentclass[
    ,final            
  ]
  {aipproc}

\layoutstyle{6x9}

\begin{document}

\title{Disentangling the EMC Effect}


\author{E. Piasetzky}{
  address={Tel Aviv University, Tel Aviv 69978, Israel}
}

\author{O. Hen}{
  address={Tel Aviv University, Tel Aviv 69978, Israel}
}

\author{L. B. Weinstein}{
  address={Old Dominion University, Norfolk, Virginia 23529, USA}
}

\begin{abstract}
The deep inelastic scattering cross section for scattering from bound nucleons differs from that of free nucleons.This phenomena, first discovered 30 years ago, is known as the EMC effect and is still not fully understood. Recent analysis of world data showed that the strength of the EMC effect is linearly correlated with the relative amount of Two-Nucleon Short Range Correlated pairs (2N-SRC) in nuclei. The latter are pairs of nucleons whose wave functions overlap, giving them large relative momentum and low center of mass momentum, where high and low is relative to the Fermi momentum of the nucleus. The observed correlation indicates that the EMC effect, like 2N-SRC pairs, is related to high momentum nucleons in the nucleus. This paper reviews previous studies of the EMC-SRC correlation and studies its robustness. It also presents a planned experiment aimed at studying the origin of this EMC-SRC correlation. 
\end{abstract}

\maketitle


\section{1. Introduction and historical overview}
The European Muon Collaboration (EMC) effect is the observation of a linear decrease in the per-nucleon Deep Inelastic Scattering (DIS) cross section ratio for nucleus A to deuterium at moderate to large four-momentum transfer, $Q^2\ge2$ (GeV/c)$^2$, and $0.3\le x_B\le0.7$ ($x_B=\frac{Q^2}{2m_N\omega}$, where $\omega$ is the energy transfer and $m_N$ is the nucleon mass). It was first observed in the DIS cross section ratio of Iron relative to Deuterium and confirmed by subsequent measurements on a wide variety of light and heavy nuclei~\cite{Aubert83,Bodek83,Gomez94,Seely09}. Ever since its discovery in 1982, the EMC effect raised a challenge to both the nuclear and particle physics communities. While extended experimental and theoretical efforts were put into studies of the origin of the EMC effect, an acceptable explanation has yet to be found. One common conclusion, rising from most modern studies, is that one can not explain the EMC effect using only traditional nuclear physics phenomena such as binding energy and nucleon momentum. This implies that the EMC effect is evidence for the existence of modification of the internal structure of bound nucleons due to the nuclear medium~\cite{Geesaman95, Norton03,Frnkfurt12,Kulagin}. For a recent review of the EMC effect see Ref.~\cite{Norton03}.

Two-Nucleon Short Range Correlations ($2N$-SRC) are states in which two nucleons have their wave function overlap in the ground state of the nuclear wave function. In these states the two nucleon system is characterized by large relative momentum and low center of mass momentum, where high and low is measured relative to the Fermi momentum of the nucleus (for medium nuclei $k_F\approx250$ MeV/c). Exclusive studies have shown that $2N$-SRC, mainly neutron-proton pairs, dominate the high momentum tail of the nuclear wave function ($p\ge k_F$)~\cite{Shneor07,Subedi08,Tang03,Piasetzky06}. Inclusive measurements at $x_B\ge1$ on a wide variety of nuclei show that in medium and heavy nuclei $2N$-SRC account for about $20-25\%$ of the nucleons in the nucleus and approximately $80\%$ of the kinetic energy carried by nucleons in the nucleus\cite{Day87,egiyan03,egiyan06}.  Under these conditions the inclusive cross section ratio for any two nuclei is constant. The scale factor of the cross section ratio, traditionally denoted as $a_2(A/d)$, is interpreted as the relative amount of high-momentum correlated nucleons in the measured nuclei~\cite{Day87,egiyan03,egiyan06,fomin12}.

Recent analysis of the EMC effect and $2N$-SRC showed that the two are correlated. It was observed that the strength of the EMC effect is strongly correlated with the relative amount of $2N$-SRC pairs in nuclei (See Fig.~\ref{fig:figure1}-A)~\cite{Weinstein11}. The observed linear correlation implies that the EMC effect, like $2N$-SRC, is related to high momentum nucleons in the nucleus. With publication of new, high precision, $2N$-SRC scaling factors~\cite{fomin12}, the EMC-SRC correlation was re-visited and its robustness was tested against different theoretical corrections used in the extractions of the $2N$-SRC scaling factors from the experimental cross section ratios~\cite{Hen12}.

 The EMC-SRC correlation was used to estimate the deuterium In Medium Correction (IMC) effect (the deviation of the DIS cross section ratio of the deuteron to a free neutron-proton pair from unity) and to extract the free neutron structure function, corrected for the IMC effect, in the $x_B$ range of $0.3$ to $0.7$~\cite{Weinstein11,Hen12}. The IMC corrected free neutron structure function was incorporated into the CTEQ-JLab (CJ) global QCD analysis and was used to constrain theoretical uncertainties in the off-shell structure of nucleons in deuterium and the d/u parton distribution ratio at the limit of $x_B\rightarrow1$~\cite{Hen11}.

In this paper we review previous studies of the EMC-SRC correlation and study the robustness of the EMC-SRC correlation, the free neutron structure function extraction and the constraints it imposes on models of the off-shell structure of deuterium. The paper is organized as follows: Sections 2 and 3 review previous studies of the EMC-SRC correlation and show that this correlation is robust against different theoretical corrections used in the extraction of $a_2(A/d)$. Section 4 presents the robustness of the previously extracted free neutron structure function, published constraints on the off-shell structure of deuterium and the d/u parton distribution ratio at the limit of $x_B\rightarrow1$. Section 5 considers the proposed connection between the EMC effect and the average nuclear binding energy. Section 6 presents a future experimental program to study medium modification of high momentum (SRC) nucleons as a possible cause of the EMC effect. In section 7 we conclude.

\begin{figure}[ht]
\begin{minipage}[b]{0.45\linewidth}
\centering
\includegraphics[width=7.5cm, height=6cm]{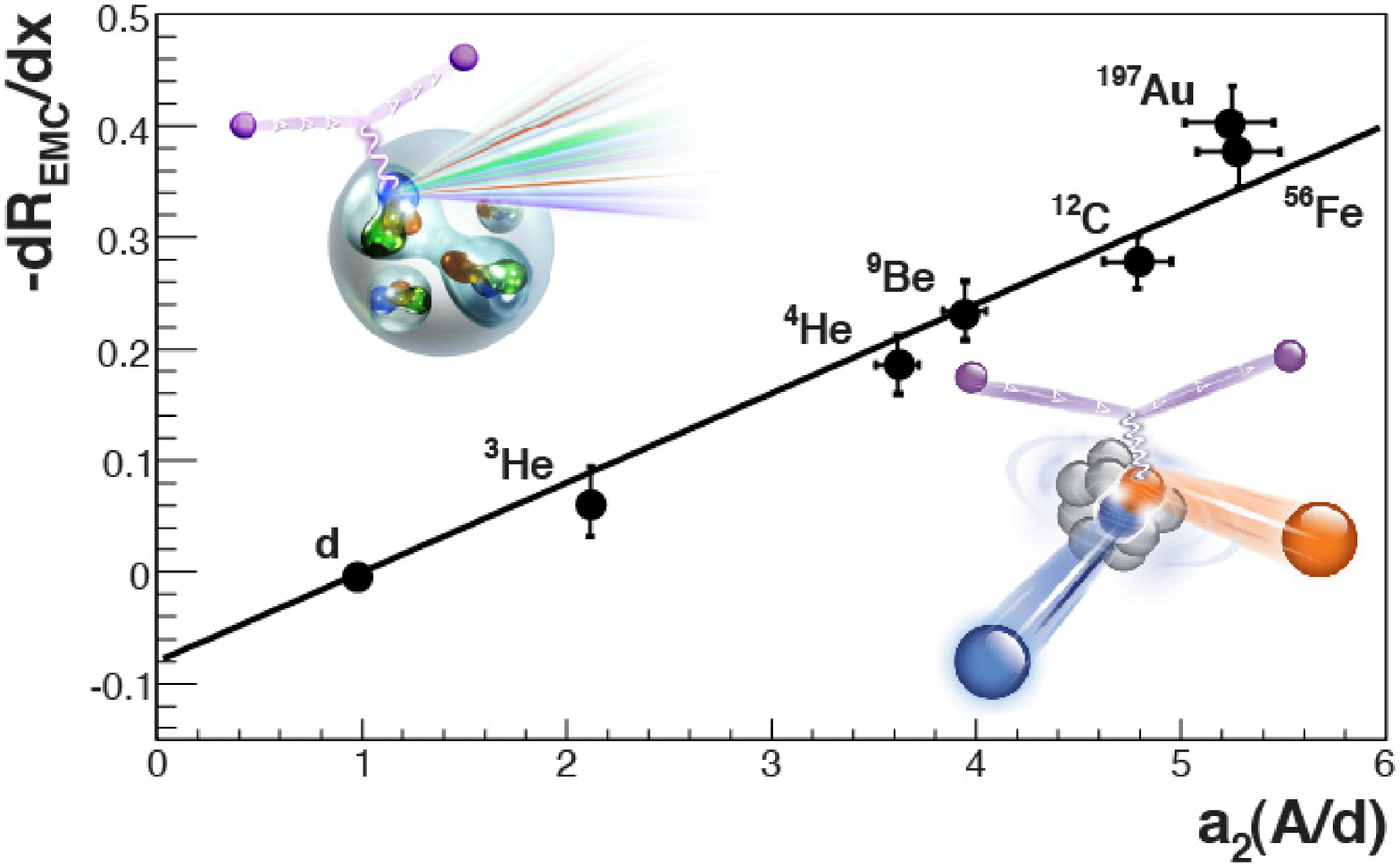}
\caption{defaultaaa}{1-A}
\label{fig:figure1}
\end{minipage}
\hspace{0.5cm}
\begin{minipage}[b]{0.45\linewidth}
\centering
\includegraphics[width=6cm, height=6cm]{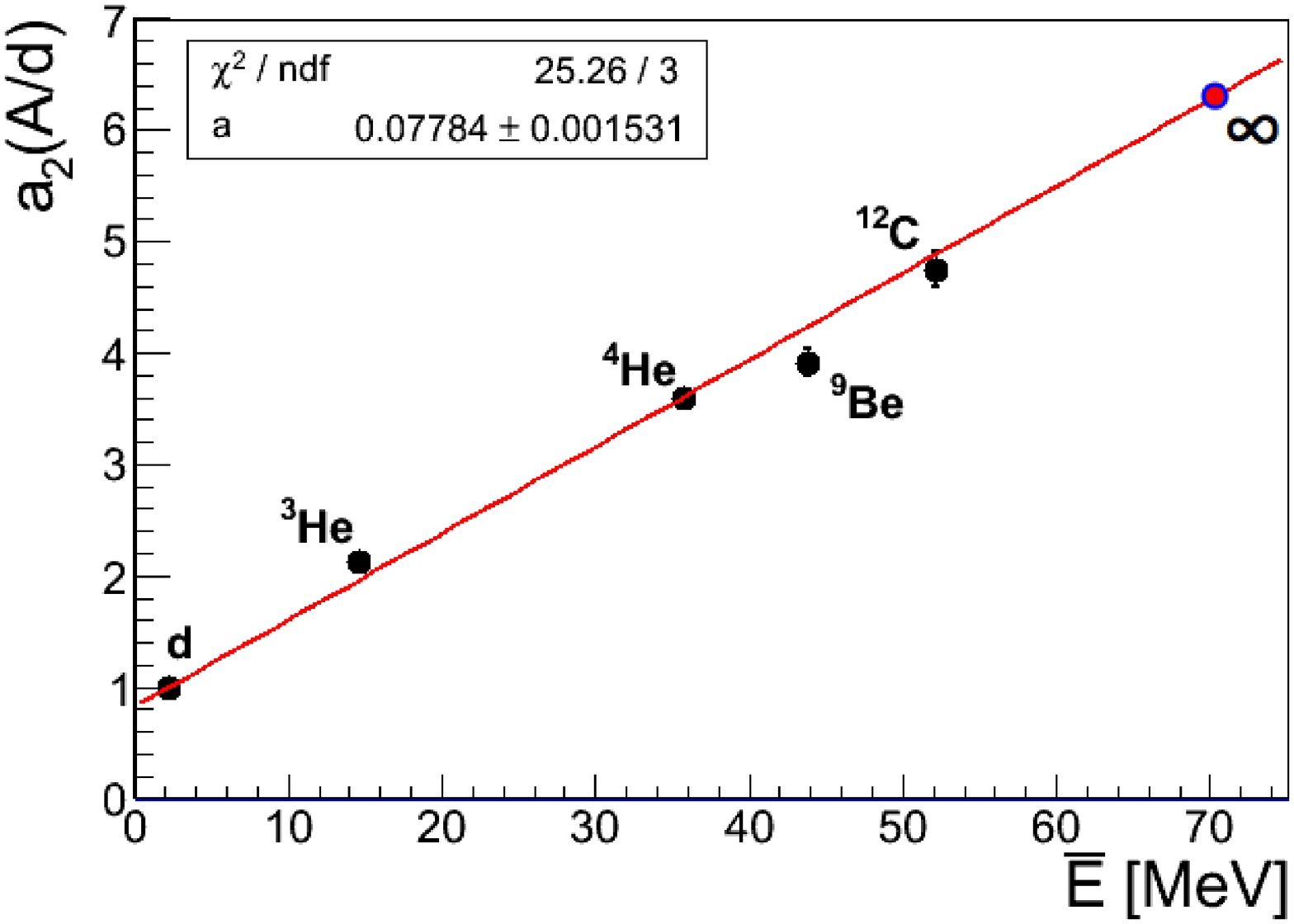}
\caption{{\bf{1-A:}} The correlation between the strength of the EMC effect and the SRC scale factors. {\bf{1-B:}} The correlation between the SRC scale factors and the average nucleon separation energy. Also shown is the prediction of $a_2(A/d)$ for isospin symmetric infinite nuclear matter. See text for details.}{1-B}
\label{fig:figure1}
\end{minipage}
\end{figure}

\section{2. Inclusive Scattering at $x_B\ge1$}
Inclusive quasi-elastic cross section ratios for nuclei A relative to deuterium, measured at large $x_B$ ($1\le x_B\le2$) and moderate four-momentum transfer ($Q^2\ge1.4$ (GeV/c)$^2$) scale~\cite{Day87,egiyan03,egiyan06,fomin12}. At these kinematics the reaction is expected to be dominated by $2N$-SRC. The scale factor of the cross section ratio, traditionally denoted as $a_2(A/d)$, is interpreted as the relative amount of high-momentum correlated nucleons in the measured nuclei~\cite{Day87,egiyan03,egiyan06,fomin12,Ciofi96}.

The extraction of the $a_2(A/d)$ scaling parameter from experimental cross section ratios requires the introduction of theoretical corrections. Different analyses applied different sets of corrections, giving rise to differences in the $a_2(A/d)$ values. This dependence of $a_2(A/d)$ on the corrections applied to the experimental data needs to be taken into account when performing a global analysis of world data.

The corrections applied to the experimental data include both traditional corrections, which do not require any a priori knowledge of $2N$-SRC pairs, and model dependent corrections which rely on theoretical models and/or previous experimental knowledge of $2N$-SRC. The traditional corrections include radiative corrections, inelastic subtractions, and coulomb corrections. The model-dependent corrections include iso-scalar corrections for asymmetric nuclei and corrections for the center-of-mass (c.m.) motion of the correlated pair. The iso-scalar correction, $((A/2)\sigma_{ep}+(A/2)\sigma_{en})/(Z\sigma_{ep}+N\sigma_{en})$, corrects the cross section to that of an imaginary nucleus with $A/2$ protons and $A/2$ neutrons.  The value of this correction is questionable since, for relative momentum between $300$ and $600$ MeV/c, $2N$-SRC are dominated by neutron-proton pairs~\cite{Shneor07,Subedi08,Tang03,Piasetzky06}. The c.m. motion correction is a correction due to the enhancement of the high momentum tail of the nuclear wave function for medium and heavy nuclei due to the c.m. motion of the pairs. The c.m. correction was claimed [16] to change the physical meaning of the $a_2(A/d)$ scaling parameter. Without this correction applied, $a_2(A/d)$ interpreted as a measure of the relative amount of high-momentum nucleons in the nucleus. With this correction applied, $a_2(A/d)$ was interpreted as a measure of the relative amount of $2N$-SRC pairs in nuclei. In order to avoid confusion, we denote $a_2(A/d)$, with the c.m. correction applied, as $R_2(A/d)$~\cite{fomin12}.

The two main analyses which extracted the SRC scaling parameters are those of Egiyan et al.~\cite{egiyan03,egiyan06} and Fomin et al.~\cite{fomin12}. Egiyan's analysis included radiative and isoscalar corrections alone. The more recent Fomin  analysis included radiative corrections, but discarded the iso-scalar corrections and instead applied coulomb, inelastic and c.m. motion corrections. The c.m. correction, which is on the order of $\sim20\%$ for medium and heavy nuclei ($A\ge12$), was based on a theoretical calculation by C. Ciofi Degli Atti et al. for Iron and an assumed A dependence~\cite{Ciofi96}. Due to the large model dependence of this correction and its unknown $A$ dependence, a $30\%$ uncertainty was applied~\cite{fomin12}.

\section{3. The EMC-SRC Correlation}
Measurements of the EMC effect on a wide variety of nuclei showed that the strength of the effect (i.e., the difference of the DIS cross section ratio of nuclei relative to deuterium from unity) increases with $A$~\cite{Gomez94,Seely09}. For many years, due to the limited experimental data available, the strength of the EMC effect seemed to scale linearly with the average nuclear density~\cite{Gomez94}. This linear dependence might indicated that the EMC effect is a property of the nuclear mean-field. However, recent high precision measurements of light nuclei showed that the strength of the effect is not proportional to the scaled nuclear density. Beryllium is the most telling outlier with an EMC effect which is anomalously large compared to its (relatively low) average density. These results seems to indicate that the EMC effect is related to the local density of the nuclear medium, with the Beryllium anomaly explained by describing it structure as two alpha clusters and an orbiting neutron, having the local density of $^4$He~\cite{Seely09}. 

If the EMC effect is indeed due to high local density fluctuations then it should be strongly correlated with $2N$-SRC, which are states of high local nuclear density. Motivated by this, Weinstein et al.~\cite{Weinstein11} performed a global analysis of world data and studied the dependence of the strength of the EMC effect on the relative amount of $2N$-SRC pairs in nuclei. There is a clear linear correlation between the strength of the EMC effect and the relative amount of $2N$-SRC pairs in  nuclei (see Fig.~\ref{fig:figure1}-A). As in previous studies, this analysis defined the strength of the EMC effect as the slope of the per nucleon DIS cross section ratio for nuclei relative to deuterium for $0.35\le x_B\le0.7$~\cite{Seely09}. The advantage of this definition is that it is insensitive to overall normalization uncertainties that could have large effect when combining data sets from different experiments.

At the time the original study Weinstein et al. was conducted, the only measurements of the SRC scaling factor $a_2(A/d)$ available were those of Frankfurt et al.~\cite{Day87} and of Egiyan et al.~\cite{egiyan03,egiyan06}. Recent data of Fomin et al.~\cite{fomin12} provided precise measurements of $a_2(A/d)$ for a wide range of nuclei, allowing us to test the robustness of the EMC-SRC correlation for different theoretical corrections to $a_2(A/d)$. A followup paper by Hen et al.~\cite{Hen12} studied the effect of these corrections to $a_2(A/d)$ on the EMC-SRC correlation. These sets of corrections included: (1) Radiative and Iso-scalar corrections as used in the Egiyan analysis, (2) Radiative, Coulomb, Inelastic, and c.m. motion corrections as used in the Fomin analysis, and (3) Radiative, Coulomb, and Inelastic corrections, as used in the Fomin analysis, but without the c.m. motion correction. The results showed that the linear EMC-SRC correlation is robust, with a linear fit giving a $\chi^2$ per degree of freedom in the range of of $0.8Ð1$ for all sets of corrections. The slope of the linear fits agree, within uncertainties, for all sets of corrections, except for the one which include the c.m. correction (i.e. Fomin analysis) which gives a slightly larger slope.

Recent work by Arrington et al.~\cite{Arrington12} tried to gain more insight on the origin of the EMC effect by comparing the quality of the EMC-SRC linear correlation without ($a_2(A/d)$) and with ($R_2(A/d)$) the c.m. motion correction. They argued that $a_2(A/d)$ is a measure of the relative amount of high momentum nucleons and is therefore a measure of large virtuality, whereas $R_2(A/d)$ is a measure of the relative amount of $2N$-SRC pairs and is therefore a measure of high local density. They also chose to scale $R_2(A/d)$ by the ratio of the total number of possible pairs in the nucleus and the number of isoscalar pairs (i.e. $N_{tot}/N_{iso} = A(A-1)/2NZ$). This scaling is presented as a correction for the ''fact'' that the experimental $R_2(A/d)$ extraction is only sensitive to neutron-proton short range pairs, while the EMC effect can be due to all short-distance nucleon-nucleon pairs, regardless of their relative momentum. 

We do not agree that the number of NN-SRC pairs should be corrected by applying this $N_{tot}/N_{iso}$ factor. The abundance of non isoscalar (pp and nn) SRC pairs was studied experimentally and theoretically and was shown to be small. The EMC effect should be related either to the mean-field nucleons or to the short range isoscalar (np) pairs but not to all possible pairs in the nucleus. 

Despite this disagreement over the $N_{tot}/N_{iso}$ factor, Arrington et al.~\cite{Arrington12} obtained very similar results to the previous study by Hen et al.,~\cite{Hen12} and showed an excellent linear correlation between the strength of the EMC effect and either scaling variable $a_2(A/d)$ or $R_2(A/d)$.

\section{4. Free Neutron Structure Function and Constrains on the $d/u$ Ratio at Large $x_B$}
The free neutron structure function is extracted by extrapolating the linear EMC-SRC correlation to the limit of $a_2(A/d)\rightarrow0$~\cite{Weinstein11}. This limit of no correlations corresponds to a free neutron-proton pair. Therefore, the y-intercept of the linear fit to the EMC-SRC correlation gives us the strength of the deuteron IMC effect (i.e., the slope of the DIS cross section ratio for a free proton-neutron pair relative to the deuteron). The shape of the EMC effect is linear for $x_B$ between $0.3$ to $0.7$ and was observed to be universal for all nuclei. Therefore, one can use the deuteron IMC slope and the known deuterium and proton structure functions to extract the free neutron structure function. Fig.~\ref{fig:figure2}-A shows the extracted ratio of the neutron to proton structure function, obtained using different $a_2(A/d)$ analyses which give different slopes to the EMC-SRC correlation. Also shown is an extraction by Arrington et al.~\cite{Arrington09}, which did not correct for any nucleon structure function modification in the deuteron. Because the SRC c.m. motion correction increases the deuteron IMC effect, we also added an extraction using a scaled down deuteron IMC effect. As can be seen, all extractions that assume the existence of a deuteron IMC affect agree, within uncertainties, and differ from the extraction of Ref.~\cite{Arrington09} at large values of $x_B$.

The IMC-corrected free neutron structure function was used to constrain theoretical uncertainties in the CTEQ-JLab (CJ) global QCD analysis~\cite{Accardi1011}. The CJ analysis extracts the patron distribution functions by correcting for off-shell effects in the deuteron in a model which depends on two parameters: (1) the average nucleon virtuality in deuterium, (2) the amount of swelling (modification of the structure functions) of the nucleons in deuterium. While the average nucleon virtuality in deuterium is well constrained by choice of wave function, the amount of swelling of the nucleon is not. We considered four deuterium wave functions (WJC-1, WJC-2, CD-Bonne, and AV18) and swelling levels of $0$ to $3\%$ and compare the resulting neutron structure function, obtained from each combination of wave function and swelling level, with the IMC corrected neutron structure function. By demanding $90\%$ confidence level on the agreement between the CJ and IMC extractions, the swelling level were constrained to $0.8\pm0.6\%$ and the d/u parton distribution ratio, at the limit of $x_B\rightarrow1$, to $0.23\pm0.09$~\cite{Hen11}. Fig.~\ref{fig:figure2}-B shows the sensitivity of the d/u ratio to the $a_2(A/d)$ analysis. As can be seen, the d/u ratio is insensitive to the choice of corrections applied in the $a_2(A/d)$ extraction.

\begin{figure}[ht]
\begin{minipage}[b]{0.45\linewidth}
\centering
\includegraphics[width=7cm, height=6cm]{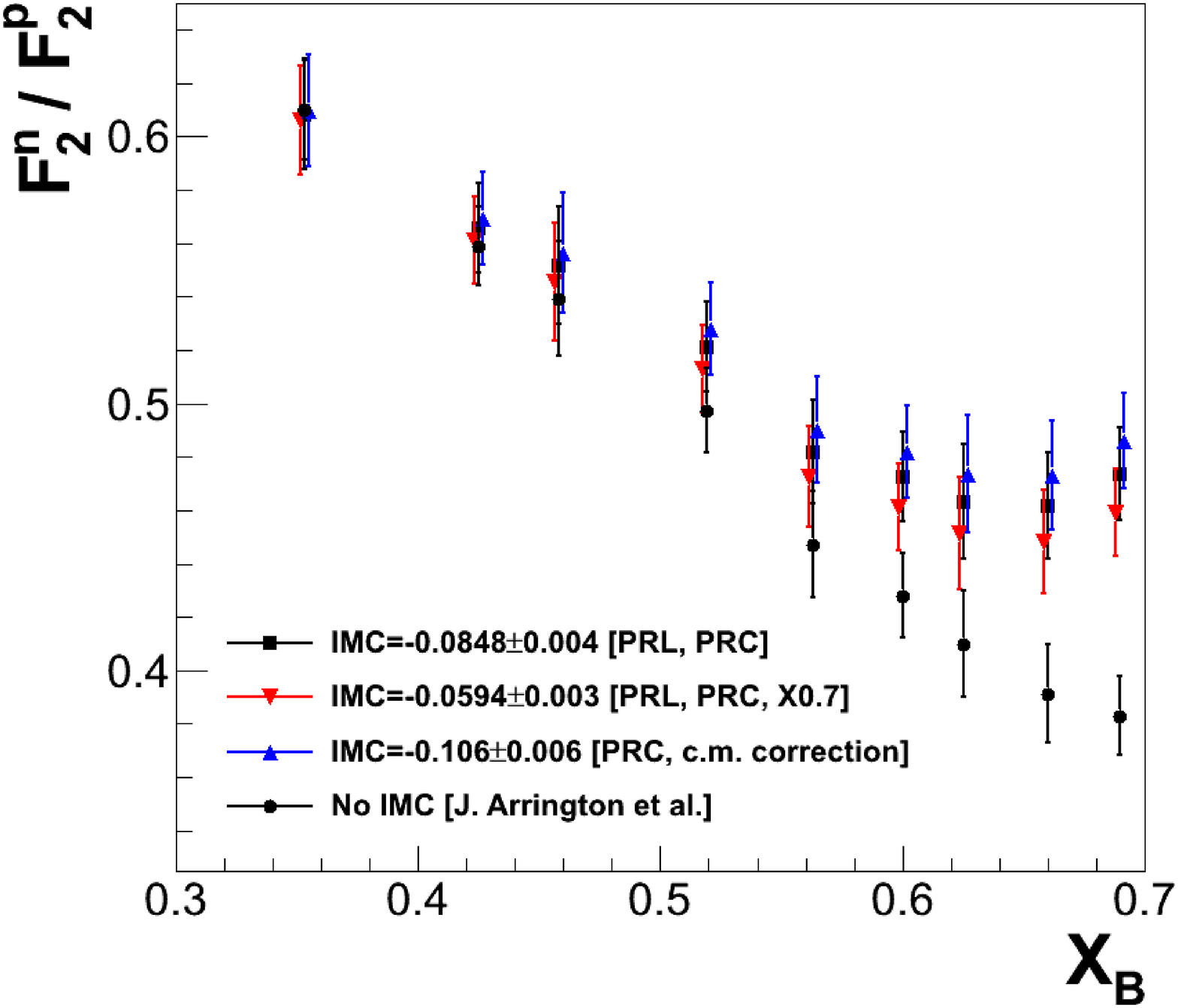}
\caption{defaultaaa}{2-A}
\label{fig:figure2}
\end{minipage}
\hspace{0.5cm}
\begin{minipage}[b]{0.45\linewidth}
\centering
\includegraphics[width=7cm, height=6cm]{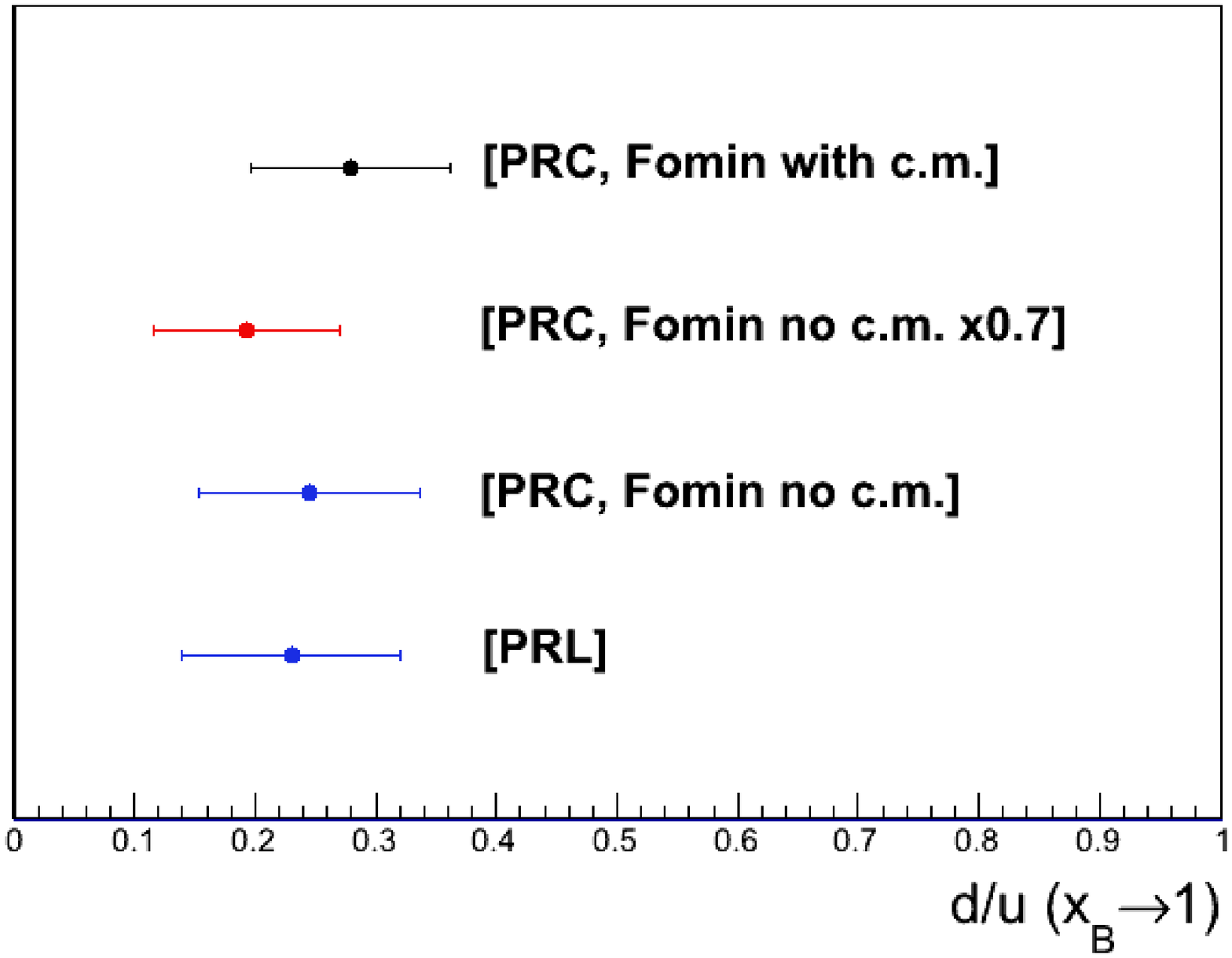}
\caption{{\bf{2-A:}} The extracted neutron to proton structured function ratio. All symbols, except for the circles, are extraction based on the EMC-SRC correlation, corrected for the deuteron IMC effect. The circles are an extraction by Arrington et al., assuming no deuteron IMC effect~\cite{Arrington09}. {\bf{2-B:}} The d/u parton distribution ratio at large $x_B$, extracted using different deuteron IMC correction factors. See text for details.}{2-B}
\label{fig:figure2}
\end{minipage}
\end{figure}

\section{5. EMC, SRC, and the Nuclear Binding Energy}
Due to their high momentum, nucleons in $2N$-SRC dominate the kinetic energy carried by nucleons in the nucleus. Neglecting 3-Body forces, the average removal energy of nucleons in the nucleus is defined as~\cite{Benhar12}:
\begin{equation} 
 \bar{E} = \bar{T}\frac{A-2}{A-1} - \frac{E_0}{A}
\end{equation} 
where A is the mass number of the nucleus, $\bar{E}$ and $\bar{T}$ are the average removal and kinetic energies of nucleons in the nucleus and $E_0$ is the nuclear binding energy obtained from nuclear masses. Recent work by O. Benhar and I. Sick calculated the average removal energy, $\bar{E}$, for light nuclei ($A\le12$) using Greens Function Monte-Carlo (GFMC) calculations of the average kinetic energy of nucleons in nuclei~\cite{Benhar12}. The obtained values were considerably larger than previous mean-field calculations and the (e,e'p) measurements of s and p shell mean removal energies, neither of which are sensitive to the effect of $2N$-SRC. Their work also showed that the strength of the EMC effect is linearly correlated with the average nucleon removal energy $\bar{E}$. We point out that since $\bar{E}$ is proportional to $\bar{T}$ which is dominated by $2N$-SRC, this is just another manifestation of the EMC-SRC correlation. This can be seen in Fig.~\ref{fig:figure1}-B which shows the correlation between $a_2(A/d)$ and $\bar{E}$. One can also calculate the average nucleon removal energy for symmetric infinite nuclear matter and use the correlation between $\bar{E}$ and $a_2(A/d)$ to determine $a_2(A/d)$ for nuclear matter (see Fig.~\ref{fig:figure1}-A). Combining this value of  $a_2(A/d)$ with the EMC-SRC correlation, one obtain an EMC effect for nuclear matter that is consistent with that of $^{56}$Fe and $^{197}$Au.

\section{6. Future experiments}
The EMC-SRC correlation suggests that the EMC effect, like $2N$-SRC, is due to high momentum nucleons in the nucleus and that the internal structure of high momentum nucleons in the nucleus is modified. To test this, experiment $E12-11-107$, approved to run as part of the $12$GeV program of Jefferson lab, will study the structure function of high momentum nucleons in deuterium~\cite{Emcsrcexpt11}. This will be done by performing DIS from nucleons in deuterium and tagging the high momentum backward recoil nucleons. At large backward recoil angles FSI are expected to be small and therefore the momentum of the recoil nucleon equals that of the nucleon the electron scattered off.

Recoil protons and neutrons will be detected in a new Large Acceptance Detector (LAD) (see Fig.~\ref{fig:figure3}-B), consisting off $132$ Time of Flight (TOF) scintillators coveting a solid angle of $1$Sr and backward angles of $90^o-180^o$ degrees at $4$ meters from the target.

The expected accuracy of the proposed measurement for the recoil protons is shown in Fig.~\ref{fig:figure3}-A. It shows the double ratio of the bound to free structure function at high and low $x_B$ ($0.6$ and $0.3$ respectively) as a function of the light-cone momentum fraction ($\alpha$) of the proton. Fig.~\ref{fig:figure3}-A also shows three theoretical models predictions (see details in~\cite{Emcsrcexpt11}). 

Furthermore, a series of approved inclusive measurements will measure the EMC effect and SRC scaling factors ($a_2(A/d)$) in a wide variety of light and heavy nuclei~\cite{arringtonexpt06,solvignonexpt11,arringtonexpt10}. These experiments will allow a detailed study of the scaling of the EMC effect and $2N$-SRC as a function different nuclear properties and as a function of each other.

\begin{figure}[ht]
\begin{minipage}[b]{0.45\linewidth}
\centering
\includegraphics[width=7cm, height=6cm]{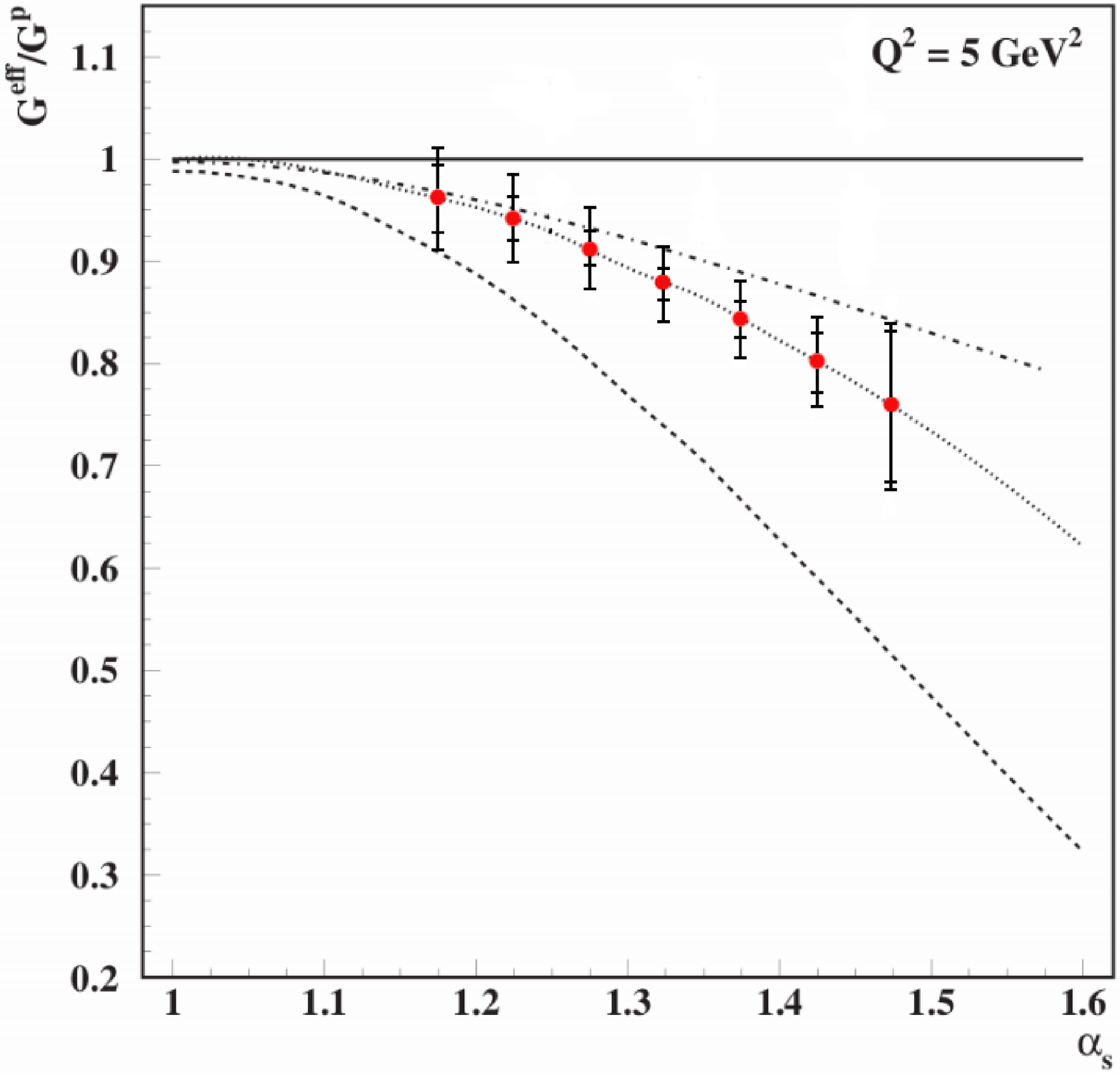}
\caption{defaultaaa}{3-A}
\label{fig:figure3}
\end{minipage}
\hspace{0.5cm}
\begin{minipage}[b]{0.45\linewidth}
\centering
\includegraphics[width=7cm, height=6cm]{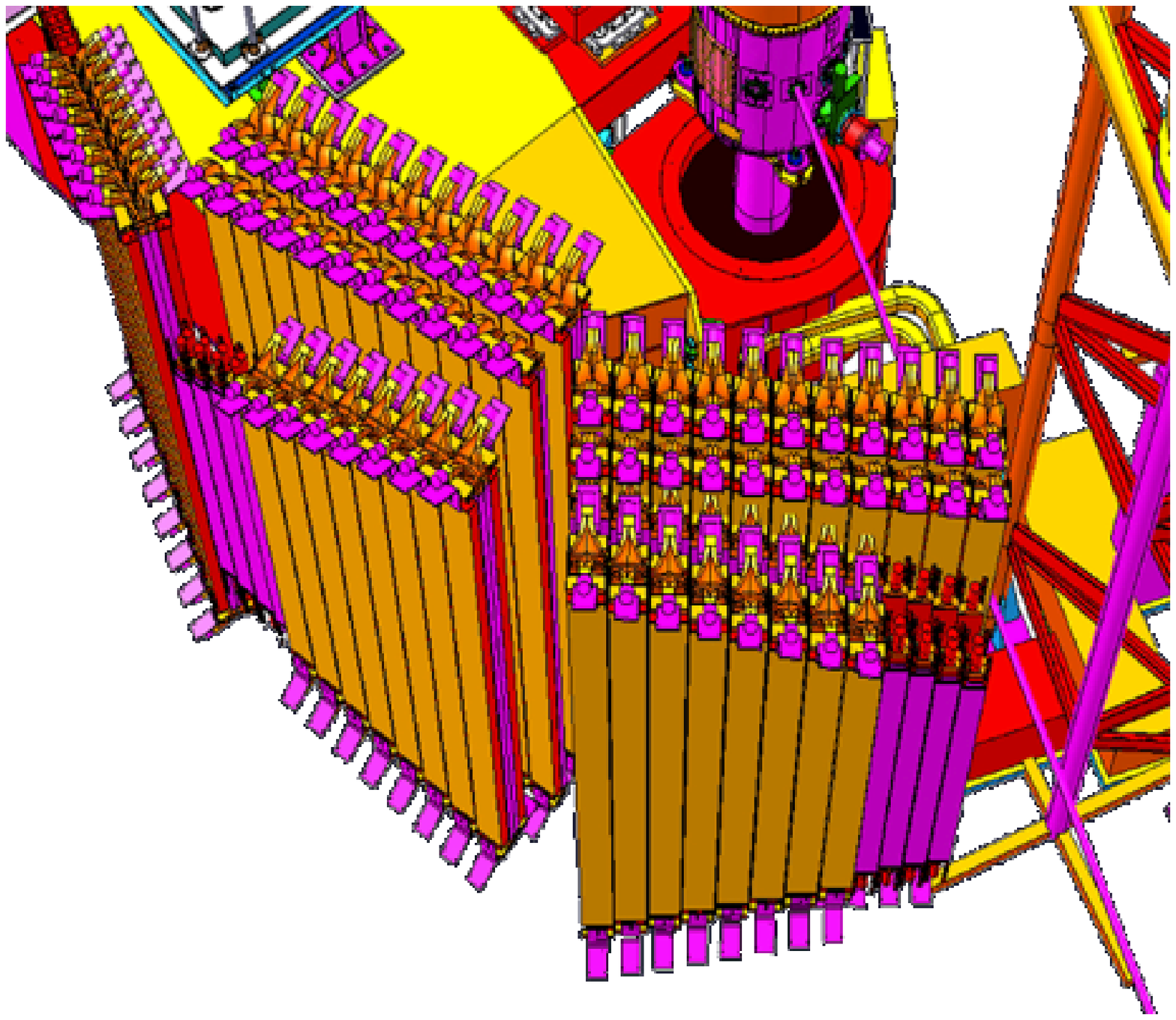}
\caption{{\bf{3-A:}} The predicted uncertainties in the measurement of the $\alpha$ dependence of the bound to free neutron structure function ratio. {\bf{3-B:}} The proposed setup of the LAD detector.}{3-B}
\label{fig:figure3}
\end{minipage}
\end{figure}

\section{7. Conclusions}
The strength of the EMC effect is correlated with the relative amount of $2N$-SRC pairs in nuclei. The correlation is robust and does not depend on the theoretical corrections applied in the extraction of $a_2(A/d)$. By extrapolating the linear fit of the EMC-SRC correlation to the limit of a free neutron-proton pair (i.e. $a_2(A/d)\rightarrow0$) one can extract the IMC effect in the deuteron and the free neutron structure function. The free neutron structure function was used to constrain theoretical uncertainties in the off-shell structure of nucleons in deuterium and the d/u parton distribution function at large $x_B$. These were also shown to be insensitive to theoretical corrections used in the extraction of $a_2(A/d)$.

The strength of the EMC effect was shown to scale as the average nucleon separation energy, extracted using GFMC calculations. The separation energy is proportional to the average kinetic energy carried by nucleons in the nucleus. The latter is dominated by $2N$-SRC, making this correlation another manifestation of the EMC-SRC correlation.

Future experiments will gain insight to the origin of the EMC effect by measuring the structure function of high-momentum nucleons.


\bibliographystyle{elsarticle-num}

\end{document}